\documentstyle[preprint,aps]{revtex}
\begin{document}
\draft

\title{Perturbative analysis for Kaplan's lattice chiral fermions}

\author{S. Aoki and H. Hirose}
\address{Institute of Physics, University of Tsukuba, Tsukuba
Ibaraki-305, Japan}

\date{\today}

\maketitle

\begin{abstract}
Perturbation theory for lattice fermions with domain wall mass terms
is developed and  is applied to
investigate the chiral Schwinger model formulated on the lattice
by Kaplan's method.
We calculate the effective action for gauge fields to one loop,
and find that it contains a longitudinal component even for anomaly-free cases.
From the effective action we obtain
gauge anomalies and Chern-Simons current without ambiguity.
We also show that the current corresponding to the fermion number
has a non-zero divergence and it flows off the wall into the extra
dimension.
Similar results are obtained for a proposal
by Shamir, who used a constant mass term with free boundaries 
instead of domain walls.
\end{abstract}
\pacs{11.15Ha, 11.30Rd, 11.90.+t}

\narrowtext
\section{Introduction}
\label{sec:int}

Construction of chiral gauge theories is one of the long-standing 
problems of lattice field theories. Because of the fermion doubling
phenomenon,
a naively discretized lattice fermion field yields $2^d$ fermion modes, half of
one chirality and half of the other, so that the theory is non-chiral.
Several lattice approaches have been proposed to define chiral gauge theories,
but so far none of them have been proven to work successfully.

Kaplan has proposed a new approach\cite{KP} to this problem.
He suggested that it may be possible to simulate the behavior of massless
chiral fermions in 2k dimensions by
a lattice theory of massive fermions in 2k+1 dimensions if
the fermion mass has a shape of a domain wall in the 
2k+1-th dimension.
He showed for the weak gauge coupling limit
that massless chiral states arise as zero-modes bound to
the 2k-dimensional domain wall while all doublers can be given large gauge 
invariant masses. If the chiral fermion content that appears 
on the domain wall
is anomalous the 2k-dimensional gauge current flows off the wall 
into the extra dimension so that the theory can not be 2k-dimensional.
Therefore he argued that
this approach possibly simulates the 2k-dimensional chiral fermions 
only for anomaly-free cases.  

His idea was tested for smooth external gauge fields.
Jansen\cite{Jan} showed numerically
that in the case of the chiral Schwinger model in 2 dimensions
with three fermions of charge 3, 4, and 5 the anomalies 
in the gauge currents cancel on the wall.
The Chern-Simons current far away from the domain wall 
was calculated in Ref.\ \cite{CS}. It is shown that
the 2k+1-th component of the current is non-zero
in the positive mass region
and zero in the negative mass region
such that the derivative of the current cancels 
the 2k-dimensional gauge anomaly on the wall, as was argued in Ref.\ \cite{KP}.

In the continuum perturbation theory 
Frolov and Slavnov\cite{FS} proposed a gauge invariant regularization
of the standard model through an infinite tower of regulator fields.
Some similarity between their proposal and the Kaplan's approach 
was pointed out by Narayanan and Neuberger\cite{Neu}.
It has been also shown that 
the chiral fermion determinant can be nonperturbatively defined as an 
overlap of two vacua\cite{Neu2}, which can be extended to lattice theories.
Using Narayanan and Neuberger's point of view, 
we observed in Ref.\ \cite{AK} that
non-gauge (chiral) anomalies are correctly reproduced within
the Frolov and Slavnov's regularization method.

The results above provide positive indications that Kaplan's method for chiral
fermion may work.
There exists, however, several potential problems in his approach.
Since the original 2k+1-dimensional  model is vector-like,
there always exists an anti-chiral mode, localized on an anti-domain wall
formed by periodicity of the extra dimension.
If the chiral mode and the anti-chiral mode are paired into
a Dirac mode, this approach fails to simulate chiral gauge theories.
Without dynamical gauge fields, the overlap between the chiral mode and 
the anti-chiral mode is suppressed as $O (e^{-L})$ where $L$ is the size of the
extra dimension. If gauge fields become dynamical, the overlap
depends on the gauge coupling. In the original paper\cite{KP}
the strong coupling limit of the gauge coupling in the extra dimension
was proposed to suppress the overlap. However,
a mean-field calculation\cite{Alt}
in this limit indicated that the chiral mode disappears and the model becomes
vector-like. 

More recently Distler and Rey\cite{Dis} pointed out that
the Kaplan's method may have a problem in reproducing fermion number
non-conservation expected in the standard model.
Using the 2-dimensional chiral Schwinger model
they argued that either the 2-dimensional fermion number current is exactly 
conserved or the light degree of freedom flows off the wall into the
extra dimension so that the model can not be 2-dimensional.

In this paper we carry out a detailed perturbative analysis of the Kaplan's
proposal for smooth background gauge fields on a finite lattice
taking the chiral Schwinger model in 2-dimensions as a concrete example.
In sect.\ \ref{sec:model},
we formulate the lattice perturbation theory for
the Kaplan's method with the periodic boundary condition.
Since translational invariance is violated by domain wall mass terms,
usual Feynman rules in the momentum space can not be used except
in the regions far away from the domain wall\cite{CS}.
To perform  perturbative calculations near or on the wall,
we use the Feynman rules in real space of the extra dimension,
as proposed in ref.\ \cite{Neu}.
We calculate the fermion propagator for the periodic boundary condition,
which reproduces the fermion propagator in ref.\cite{Neu}
near the origin of the extra dimension.
A similar calculation is also made for the constant fermion mass
with {\it free} boundaries in the 2k+1-th dimension.
As shown by Shamir\cite{Sham} the constant mass term 
with this boundary condition
can also produce the chiral zero mode on the 2k-dimensional 
boundary.
The results are similar  but simpler than those by the Kaplan's method.
In sect.\ \ref{sec:2dim},
using the Feynman rules of sect.\ \ref{sec:model}
we calculate a fermion one-loop effective action for the U(1) gauge field
of the chiral Schwinger model simulated by the Kaplan's method.
We find that the effective action
contains the longitudinal component as well as
parity-odd terms,
and that this longitudinal component, which
breaks gauge invariance, remains non-zero even for anomaly-free cases.
This result is compared with those of the conventional Wilson fermion 
formulation of this model\cite{chiral}.
In sect.\ \ref{sec:anomaly}
we derive gauge anomalies as well as the Chern-Simons current
from the effective action without ambiguity.
Then we show that the current corresponding to
the fermion number has a non-zero divergence 
and the fermion number current flows off the walls into the extra dimension.
In sect.\ \ref{sec:concl}, we give our conclusions and discussions.

\section{Action, Fermion Propagator and Chiral Zero Modes}
\label{sec:model}

\subsection{Lattice Action}
We consider a vector gauge theory in D=2k+1 dimensions with
a domain wall mass term.  For later convenience we use the
notation of ref.\cite{Neu}, where the fermionic action is written in terms of
a d=2k dimensional theory with infinitely many flavors.
Our action is denoted as
\begin{equation}
S= S_G + S_F .
\end{equation}
The action for gauge field $S_G$ is given by
\begin{eqnarray}
S_G & =& \beta \sum_{n,\mu>\nu}\sum_s {\rm Re}\{ {\rm Tr} [U_{\mu\nu}(n,s)]\}
	\nonumber \\
& +& \beta_D \sum_{n,\mu}\sum_s {\rm Re} \{ {\rm Tr} [U_{\mu D}(n,s)]\}
\end{eqnarray}
where $\mu$, $\nu$ run from 1 to $d$, 
$n$ is a point on a d-dimensional lattice and $s$  a coordinate in the
extra dimension,
$\beta$ is the inverse gauge coupling for plaquettes $U_{\mu\nu}$
and $\beta_D$  that for plaquettes $U_{\mu D}$.
In general we can take $\beta \not= \beta_D$.
The fermionic part of the action $S_F$ is given by
\widetext
\begin{eqnarray}
S_F & = & {1\over 2}\sum_{n,\mu}\sum_s  \bar\psi_s(n)\gamma_\mu
[U_{s,\mu}(n)\psi_s(n+\mu ) - U^\dagger_{s,\mu}(n-\mu)\psi_s(n-\mu ) ] 
	\nonumber \\
 &+&  \sum_n \sum_{s,t} \bar\psi_s(n) [ M_0 P_R + M_0^\dagger P_L]_{st}
\psi_t(n) 
\nonumber \\
 & + & {1\over 2}\sum_{n,\mu}\sum_s  \bar\psi_s(n)
[U_{s,\mu}(n)\psi_s(n+\mu ) + U^\dagger_{s,\mu}(n-\mu)\psi_s(n-\mu ) 
    -2\psi_s(n) ] 
\label{actionf}
\end{eqnarray}
where $s$, $t$ are considered as flavor indices, 
$P_{R/L} = (1 \pm \gamma_{2k+1})/2$,
\narrowtext
\begin{mathletters}
\begin{eqnarray}
(M_0)_{st} & = & U_{s,D}(n)\delta_{s+1,t}- a(s)\delta_{st} 
	\\
(M_0^\dagger)_{st} & = & U^\dagger_{s-1,D}(n)\delta_{s-1,t}- a(s)\delta_{st} ,
\end{eqnarray}
\end{mathletters}
and $U_{s,\mu}(n)$, $U_{s,D}(n)$ are link variables for gauge fields.
We consider the above model with periodic boundaries
in the extra dimension,
so that $s$, $t$ run from $-L$ to $L-1$, and we take
\begin{equation}
a(s) = 1 - m_0[ {\rm sign}(s+{1\over 2}) \cdot {\rm sign}(L-s-
{1\over 2})]
= \left\{ 
\begin{array}{ll}
1-m_0, &  -{1\over 2} < s < L-{1\over 2} \\
1+m_0, & -L-{1\over 2} < s < -{1\over 2}
\end{array}
\right.
\end{equation}
for $-L \leq s < L $.  It is easy to see\cite{Neu} that  $S_F$ above is 
identical to the Kaplan's action in D=2k+1 dimensions\cite{KP}
with the Wilson parameter $r=1$. In fact the second term 
in eq.\ (\ref{actionf}) can be rewritten as
\begin{eqnarray}
& {1\over 2}& \bar\psi_s\gamma_D [U_{s,D}\psi_{s+1}-U_{s-1,D}\psi_{s-1}] 
	\nonumber \\
&+& {1\over 2}\bar\psi_s[U_{s,D}\psi_{s+1}+U_{s-1,D}\psi_{s-1}
-2\psi_s] + M(s) \bar\psi_s \psi_s  
\end{eqnarray}
with $M(s)=m_0 [ {\rm sign}(s+1/2) \cdot {\rm sign}(L-s-1/2)]$.
Note that our action is slightly different from that of ref.\cite{Neu}:
we have the D-th component of the link variable $U_{s,D}(n)$ 
and all link variables have $s$ dependence. With the gauge fixing condition
$U_{s,D}(n) =1$ for all $s$ and $n$\cite{Dis},
our action becomes almost identical to that of ref.\cite{Neu}, but
still the $s$ dependence exists in our link variables in d dimensions.
The model in ref.\cite{Neu} corresponds to our model at 
$\beta_D =\infty$, where $s$ dependences of gauge fields are
completely  suppressed, and the model at $\beta_D = 0$  was
investigated by the mean field method\cite{Alt}.

\subsection{Chiral Zero Modes}
We now consider chiral zero modes of the action $S_F$ in
the weak coupling limit , i.e. $\forall U_{s,\mu} = \forall U_{s,D}=1$.
According to ref.\cite{Neu}, the right-handed zero modes are given
by  zero modes of the operator $M$ and the left-handed zero modes 
by  those of the operator $M^\dagger$, where
\begin{equation}
(M)_{st} = (M_0)_{st} + {\nabla (p)\over 2}\delta_{st}, \
(M^\dagger )_{st} = (M_0^\dagger )_{st} + {\nabla (p)\over 2}\delta_{st}
\end{equation}
with $\nabla (p)\equiv \sum_{\mu=1}^d 2 [\cos (p_\mu a) - 1 ]$
in momentum space of d dimensions.
It is noted that $ 0 \leq -\nabla(p) \leq 4d$ and zero modes exist
if and only if $ -\nabla (p) \le 2m_0$\cite{CS}.
Hereafter we only consider the case that $ 0 < m_0 < 2$.
In this range of $m_0$, there is only one right-handed zero mode $u_R$
satisfying $ M \cdot u_R = 0$, which is given by
\begin{equation}
u_R(s)  = \left\{ \begin{array}{ll}
\displaystyle (1-\nabla (p)/2-m_0)^s C_0^{-1}  & \mbox{for $s \geq 0$} \\
	& \\
\displaystyle (1-\nabla (p)/2+m_0)^s C_0^{-1}  & \mbox{for $s <  0$}
\end{array} \right.
\end{equation}
where the d-dimensional momentum $p$ has to be restricted to
$ 0 \leq m_0+\nabla (p) /2 $ so that
$(1-\nabla/ 2 -m_0) \leq 1$. The normalization constant $C_0$ takes the value
\begin{equation}
{1-(1-\nabla (p)/2-m_0)^L\over m_0+\nabla (p)/2} +
{1-(1-\nabla (p)/2+m_0)^{-L}\over m_0 - \nabla (p)/2}  .
\end{equation}
This zero mode is localized around $s=0$.
On a finite lattice (i.e. $L\not= \infty$) with the periodic boundary 
condition, there exists another zero mode $u_L$ with the opposite
chirality satisfying $ M\cdot u_L = 0 $, which is given by
$ u_L(s) = u_R(L-t-1)$ and is localized around $s=L$.
The overlap between the two zero modes vanishes exponentially as 
$L\rightarrow \infty$;
\begin{eqnarray}
\sum_{s=-L}^{L-1} u_R(s) u_L(s) & =& C_0^{-2} L\times
 (1-{\nabla (p)\over2}-m_0)^L 
	\nonumber \\
&\times &
(1-{\nabla (p)\over 2}+m_0)^{-L} 
\longrightarrow 0, 
\end{eqnarray}
We illustrate the shape of the two zero modes $u_R$ and $u_L$ 
at $m_0= 0.1$ and 0.5 for $ p =0$ in Fig.\ \ref{zero}.

\subsection{Fermion Propagator and Zero Modes}
The fermion propagator in d-dimensional momentum space and
in real D-th space has been obtained in ref.\cite{Neu} for
the infinite D-th space(i.e. $L=\infty$ ). 
It is not difficult to obtain
the fermion propagator for a finite lattice with
periodic boundaries.  We have
\begin{eqnarray}
S_F(p)_{st} & = &- \left[ [(i\sum_\mu \gamma_\mu \bar p_\mu+ M ) G_L (p)]_{st}
P_L \right.
	\nonumber \\
& + &
\left. [(i\sum_\mu \gamma_\mu \bar p_\mu+ M^\dagger ) G_R (p)]_{st}P_R \right]
\label{fprop}
\end{eqnarray}
where
\begin{equation}
G_L (p) ={1\over  \bar p^2 + M^\dagger M },
\quad 
G_R (p) ={1\over  \bar p^2 + M M^\dagger }
\end{equation}
with $\bar p_\mu = \sin (p_\mu a)$ .
Explicit expressions for $G_L$ and $G_R$ are complicated in general,
but they become simple for large $L$ where we  neglect terms of order
$O(e^{-cL})$ with $c > 0$.  
We obtain
\widetext
\begin{equation}
G_L(p)_{st}   = 
\left\{ \begin{array}{ll}
B e^{-\alpha_+|s-t|} + (A_L-B) e^{-\alpha_+(s+t)} + 
(A_R-B) e^{-\alpha_+(2L-s-t)},
& (s,t \ge 0) \\
	&       \\
A_Le^{-\alpha_+s+\alpha_-t} +A_Re^{-\alpha_+(L-s)-\alpha_-(L+t)},
& (s\ge 0,\ t\le 0) \\
	&              \\
A_Le^{\alpha_-s-\alpha_+t} + A_Re^{-\alpha_-(L+s)-\alpha_+(L-t)},
&  (s\le 0,\ t\ge 0) \\
 	&              \\
C e^{-\alpha_-|s-t|}+ (A_L-C) e^{\alpha_-(s+t)} + 
(A_R-C) e^{-\alpha_-(2L+s+t)},
& (s,t \le 0) 
\end{array}  \right. 
\label{gl}
\end{equation}
\begin{equation}
G_R(p)_{st}  = 
\left\{ \begin{array}{ll}
B e^{-\alpha_+|s-t|}  + (A_R-B) e^{-\alpha_+(s+t+2)} + 
(A_L-B) e^{-\alpha_+(2L-s-t-2)},
& (s,t \ge -1) \\
	&               \\
A_Re^{-\alpha_+(s+1)+\alpha_-(t+1)} +A_Le^{-\alpha_+(L-s-1)-\alpha_-(L+t+1)},
& (s\ge -1,\  t\le -1) \\
 	&              \\
A_Re^{\alpha_-(s+1)-\alpha_+(t+1)} +A_Le^{-\alpha_-(L+s+1)-\alpha_+(L-t-1)},
& (s\le -1,\  t\ge -1) \\
 	&              \\
C e^{-\alpha_-|s-t|} +  (A_R-C) e^{\alpha_-(s+t+2)} + 
(A_L-C) e^{-\alpha_-(2L+s+t+2)},
& (s,t \le -1) 
\end{array}  \right. 
\label{gr}
\end{equation}
\narrowtext where
\begin{mathletters}
\begin{equation}
a_{\pm} = 1 -{\nabla (p)\over 2}\mp m_0
\end{equation}
\begin{equation}
\alpha_{\pm} =  {\rm arccosh} [{1\over 2}(a_{\pm}+{1+\bar p^2\over a_{\pm}})]
\geq 0, 
\end{equation}
\begin{equation}
A_L  = {1\over a_+  e^{\alpha_+} - a_-e^{-\alpha_-}}, \qquad
A_R ={1\over a_-  e^{\alpha_-}- a_+e^{-\alpha_+}} 
\end{equation}
\begin{equation}
B = {1\over 2 a_+ {\rm sinh} \alpha_+}, \qquad
C = {1\over 2 a_- {\rm sinh} \alpha_-} . 
\end{equation}
\end{mathletters}

For $|s|, |t| \ll L$ the propagator above coincides with  
that of ref.\cite{Neu}.  
From the form of $A_L$, $A_R$, $B$ and $C$,
it is easy to see\cite{Neu}
that singularities occur only in $A_L$ at $p=0$;
\begin{equation}
A_L \longrightarrow {m_0(4-m_0^2) \over 4 p^2 a^2 },
\qquad p\rightarrow 0  .
\end{equation}
Therefore the propagator $G_L$ describes a massless right-handed
fermion around $s,t = 0$ and $G_R$ a massless left-handed fermion
around $|s|,|t| = L$, which correspond to the two zero modes in the
previous subsection.
Later we use the above forms of the fermion propagator
to calculate fermion one-loop diagrams.
It is also noted that the fermion propagator away from the two domain walls
approaches the Wilson fermion propagator 
with a {\it constant} mass term, i.e. ,
\begin{eqnarray}
S_F(p) & \rightarrow
&  \int {dp_D\over 2\pi} 
{e^{ia p_D(s-t)} \over i\gamma\cdot p + i\gamma_D p_D  \pm m_0
-\nabla (p) +1-\cos (p_D a)}
\label{wil}
\end{eqnarray}
for $1 \ll |s|, |t|, |L-s|, |L-t|$ with $s\cdot t > 0$, where $+m_0$ is taken
for  $s,t > 0$ and $-m_0$ for $s,t < 0$.
Therefore the calculation in ref.\cite{CS} is valid in this region
of $s$ and $t$.   

Before closing this subsection, we give the form of propagator for the Shamir's
free boundary fermions\cite{Sham}.
This is again given by eq.(\ref{fprop})
with $M_{st}=\delta_{s+1,t}-a_+ \delta_{s,t}$ 
and $M^\dagger_{st}=\delta_{s-1,t}-a_+ \delta_{s,t}$.
For large $L$, it becomes\cite{Sham}  
\begin{equation}
G_L(p)_{st}  =  
B e^{-\alpha_+|s-t|} + A'_L e^{-\alpha_+(s+t)} 
+A'_Re^{\alpha_+(s+t-2L)} 
\label{shl}
\end{equation}
\begin{equation}
G_R(p)_{st}  =  
B e^{-\alpha_+|s-t|}+ A'_R e^{-\alpha_+(s+t)} + 
A'_Le^{\alpha_+(s+t-2L)} ,
\label{shr}
\end{equation}
where 
\begin{equation}
A'_L = B {1-a_+e^{-\alpha_+}\over a_+e^{\alpha_+}-1}, \qquad
A'_R = - Be^{-2\alpha_+}
\end{equation}
and now $ 0 \le s,t \le L$.
Singularities occur only in $A'_L$ at $p=0$ such that
\begin{equation}
A'_L \rightarrow {m_0(2-m_0)\over p^2 a^2} .
\end{equation}
Thus, the propagator describes a right-handed massless
fermion around one boundary at $s,t = 0$ and 
a left-handed massless around the other boundary at $s,t = L$.

\subsection{Fermion Feynman Rules }
In this subsection, we write down the lattice Feynman rules for fermions
relevant for fermion one-loop calculations, 
which will be performed in the next section.
We first choose the axial gauge fixing $U_{s,D}=1$.
Although the full gauge symmetries in D dimensions are lost,
the theory is still invariant under gauge 
transformations independent of $s$ \cite{Dis}.  
We consider the limit of small d-dimensional gauge coupling, and take
\begin{equation}
U_{s,\mu}(n) = exp[i a g A_\mu (s, n+\mu/2)]
\end{equation}
where $a$ is the lattice spacing, and
$g \propto 1/\sqrt\beta$  
is the gauge coupling constant whose mass dimension is $2-D/2$
(mass dimension of the gauge fields $A_\mu$ is $D/2-1$).
It is noted that the other gauge coupling $g_s \propto 1/\sqrt{\beta_s}$
is not necessarily small and can be made arbitrary large.
We consider Feynman rules in momentum space for the physical
d dimensions but in real space for the extra dimension.
\begin{itemize}
\item The fermion propagator is given by
\begin{equation}
< \psi_s (-p) \bar\psi_t (p) > =  S_F(p)_{st}.
\end{equation}
where $S_F$ has been given in eq.(\ref{fprop}) with
eqs.(\ref{gl},\ref{gr}) for the Kaplan's fermions or with
eqs.(\ref{shl},\ref{shr}) for the Shamir's fermions.

\item The fermion vertex coupled to a single gauge field is given by
\begin{equation}
ag \bar\psi_s(q) \partial_\mu [ S_F^{-1}({q+p\over 2})]_{ss}
A_\mu(s,p-q)\psi_s(-p).
\end{equation}
Here 
$\partial_\mu S_F^{-1}(q) = \displaystyle {\partial S_F^{-1}(q)\over 
\partial (q_\mu a)} = iC_\mu(q)\gamma_\mu + S_\mu (q)$ with
$C_\mu (q)= \cos (q_\mu a)$ and $S_\mu (q)= \sin (q_\mu a)$.
From this form of the vertex it is easy to see that
the fermion tadpole diagram for an external gauge field vanishes identically.
\item The fermion vertex with two gauge fields is given by
\begin{equation}
- a^2 {g^2\over 2} 
\bar\psi_s(q) [\partial^2_\mu S_F^{-1}({q+p\over 2})]_{ss}
A^2_\mu(s,p-q)\psi_s(-p)
\end{equation}
where $A_\mu^2(s,p) = A_\mu (s,p-p_1) A_\mu(s,p_1)$ with $p$ and $p_1$ fixed.
\end{itemize}

\section{Perturbative Calculations for the Chiral Schwinger Model}
\label{sec:2dim}
In the following two sections 
we analyze the chiral Schwinger model
formulated via the Kaplan's method for lattice chiral fermions.
Using the Feynman rules of the previous section
for $D=3$,
we calculate the effective action for external gauge fields,
from which we
derive gauge anomalies, Chern-Simons current, and anomaly of the
fermion number current. 
We perform the calculations for the Shamir's method in parallel
with those for the Kaplan's method.

\subsection{Effective Action at Fermion One-Loop}
Since $ [ A_\mu, A_\nu ]=0$ for U(1) gauge fields, all diagrams with 
odd number of external gauge fields vanishes identically.  
Furthermore diagrams with four or more external gauge fields
are all convergent. Therefore only the diagrams with two external gauge
fields are potentially divergent. The effective action
for two external gauge fields is denoted by
\begin{eqnarray}
S_{eff}^{(2)} & \equiv & -{g^2\over 2}\sum_{p,s,t}
A_\mu (s,p) A_\nu (t,-p) I^{\mu\nu}(p)_{st}
	\nonumber \\
&=& -{g^2\over 2}\sum_{p,s,t} A_\mu (s,p) A_\nu (t,-p)
 [I_a^{(2)} + I_b^{(2)}]^{\mu\nu} _{st},
\end{eqnarray}
where 
\begin{eqnarray}
[I_a^{(2)}]^{\mu\nu} _{st}  & = & 
\int_{-\pi/a}^{\pi/a} {d^2 q\over (2\pi)^2}
{\rm tr} \left\{[\partial_\mu S_F^{-1}(q+{p\over 2})\cdot S_F(q+p)]_{st}
	\right.        \nonumber \\
& \times & \left.
[\partial_\nu S_F^{-1}(q+{p\over 2})\cdot S_F(q)]_{ts}\right\}\times a^2
\end{eqnarray}
and 
\begin{equation}
[I_b^{(2)}]^{\mu\nu} _{st} = -\delta_{st} \delta_{\mu\nu} 
\int_{-\pi/a}^{\pi/a} {d^2 q\over (2\pi)^2}
{\rm tr} [\partial_\mu^2 S_F^{-1}(q)\cdot S_F(q)]_{ss} \times a^2 
\end{equation}
with tr meaning  trace over spinor indices.

\subsection{Evaluation of Zero Mode Contributions}
To evaluate  $I^{\mu\nu}(p)$ we  decompose it into two parts as
\begin{equation}
I^{\mu\nu}(p) = I^{\mu\nu}_0(p)+ [I^{\mu\nu}(p)-I^{\mu\nu}_0(p)]
\end{equation}
where $I^{\mu\nu}_0$ is the contribution of zero modes and
$I^{\mu\nu}-I_0^{\mu\nu}$ is the remaining contribution.
For $I_0^{\mu\nu}$ we replace the integrand of $I^{\mu\nu}$
with that in the $a\rightarrow 0$ limit, and we obtain
\begin{eqnarray}
I^{\mu\nu}_0(p)_{st} & = &
\sum_X \int_{-\nabla (q)\le 2m_0} {d^2 q\over (2\pi)^2} 
	\nonumber \\
& \times &
{\rm tr} \left[ 
i\gamma_\mu (-i\gamma_\alpha (q+p)_\alpha a ) G_X^0(q+p)_{st} P_X
	\right.        \nonumber \\
& \times &
\left.
i\gamma_\nu (-i\gamma_\beta (q+p)_\beta a ) G_X^0(q+p)_{ts} P_X]  
\right] \times a^2 ,
\end{eqnarray}
with $X=L$ for $|s|,|t| \approx 0$, or $X=R$ for $|s|,|t| \approx L$.
The zero mode propagators $G_X^0$ are given by
\begin{equation}
G_X^0(q)_{st} = \lim_{a\rightarrow 0} G_X(q)_{st}
= {1\over q^2 a^2 } F_X (s,t)
\end{equation}
where $ F_X(s,t) = F_X(t,s)$ and
\widetext
\begin{equation}
F_L(s,t)  =  {m_0(4-m_0^2)\over 4 } \times \left\{
\begin{array} {ll}
 (1- m_0)^{s+t}  & \mbox{\ for $s,t \ge 0$ } \\
	& \\
 (1- m_0)^s(1+m_0)^t & \mbox{\ for $s \ge 0$ and $t < 0$ } \\
	& \\
 (1+m_0) ^{s+t}  & \mbox{\ for $s,t \le 0$ } 
\end{array} \right.
\end{equation}
\begin{equation}
F_R(s,t) = {m_0(4-m_0^2)\over 4 } \times \left\{
\begin{array} {ll}
 (1- m_0)^{2L-s-t-2}  & \mbox{\ for $s,t \ge 0$ } \\
	& \\
 (1- m_0)^{L-s-1} (1+m_0)^{-L-t-1} & \mbox{\ for $s \ge 0$ and $t < 0$ } \\
	& \\
 (1+m_0) ^{-2L-s-t-2}  & \mbox{\ for $s,t \le 0$ }
\end{array} \right.
\end{equation}
\narrowtext for the Kaplan's fermion with the domain wall mass terms, and
\begin{mathletters}
\begin{eqnarray}
F_L(s,t) &=& m_0 (2-m_0) (1-m_0)^{s+t}
	\\
F_R(s,t) &=& m_0 (2-m_0) (1-m_0)^{2L-s-t}
\end{eqnarray}
\end{mathletters}
for the Shamir's fermion with the constant mass terms and free boundaries.

We evaluate $I^{\mu\nu}_0(p)$ in the $a\rightarrow 0$ limit.
In this limit
\begin{eqnarray}
\displaystyle
&\displaystyle \lim_{a\rightarrow 0}& \int_{-\nabla (q)\le 2m_0} 
{d^2 q\over (2\pi)^2}
{\rm tr}[P_X\gamma_\mu\gamma_\alpha\gamma_\nu\gamma_\beta ]
{(q+p)_\alpha q_\beta \over (q+p)^2 q^2}
	\nonumber \\
& = & \displaystyle
 \int_{-\infty}^{\infty} {d^2 q\over (2\pi)^2}
{\rm tr}[P_X\gamma_\mu\gamma_\alpha\gamma_\nu\gamma_\beta ]
{(q+p)_\alpha q_\beta \over (q+p)^2 q^2}
        \nonumber \\
& = & \displaystyle
{1\over 2\pi}\left[ \delta_X 
i\epsilon^{\mu\alpha}{p_\nu p_\mu\over p^2} +
(\delta^{\mu\nu}-{p_\mu p_\nu\over p^2})-{\delta^{\mu\nu}\over 2}\right] ,
\end{eqnarray}
therefore we obtain
\begin{eqnarray}
\lim_{a\rightarrow 0} I^{\mu\nu}_0(p) & = &
\sum_X {1\over 2\pi}[\delta_X i\epsilon^{\mu\alpha}{p_\nu p_\mu\over p^2} +
(\delta^{\mu\nu}-{p_\mu p_\nu\over p^2})-{\delta^{\mu\nu}\over 2}]
        \nonumber \\
&\times & F_X(s,t)^2
\end{eqnarray}
where $\delta_L=1$ and $\delta_R=-1$.
It is noted that  $F_X$ satisfies
\begin{equation}
\sum_{t}F_X(s,t)^2 = F_X(s,s) , \qquad
\sum_{s,t}F_X(s,t)^2 =1  .
\label{sumf}
\end{equation}

\subsection{Evaluation of Remaining Contributions}
We consider the remaining terms in $I^{\mu\nu}$.
Since the combination $I^{\mu\nu}(p) - I^{\mu\nu}_0(p)$ is infra-red finite,
we can change the integration variable from $q$ to $ q a$ 
and take the $a\rightarrow 0$ limit in the integrand. Thus we obtain
\begin{equation}
\lim_{a\rightarrow 0} I^{\mu\nu}(p) - I^{\mu\nu}_0(p)
 =I^{\mu\nu} (0) = {1\over 2\pi}\cdot
[i \epsilon^{\mu\nu}\Gamma_{CS} +\delta^{\mu\nu} K]
\end{equation}
where
\begin{eqnarray}
\Gamma_{CS}(s,t)& = & {\epsilon^{\mu\nu}\over i}
2\pi 
\int {d^2q\over (2\pi)^2} {\rm tr}\left\{ [\partial_\mu S_F^{-1}(q)S_F(q)]_{st}
	\right.	\nonumber \\
& \times & \left. 
[\partial_\nu S_F^{-1}(q)S_F(q)]_{ts}\right\}
\label{gcs}
\end{eqnarray}
and
\begin{eqnarray}
& \displaystyle K (s,t)  =  2\pi
\int {d^2q\over (2\pi)^2}\left[
{\rm tr}\left\{ [\partial_\mu S_F^{-1}(q)S_F(q)]_{st}
	\right. \right. \nonumber \\
& \times  \left. \left.
[\partial_\mu S_F^{-1}(q)S_F(q)]_{ts}\right\}
- \delta_{st}{\rm tr} [\partial_\mu^2 S_F^{-1}(q)S_F(q)]_{ss} \right]  .
\end{eqnarray}
Here no summation over $\mu$, $\nu$ is implied.

The parity-odd term $\Gamma_{CS}$ is the coefficient function of 
a 3-dimensional Chern-Simons term in the axial gauge\cite{Dis}, 
which satisfies $\Gamma_{CS}(s,t)= - \Gamma_{CS}(t,s)$.
It is easy to show that
\begin{eqnarray}
& \displaystyle \sum_t  \Gamma_{CS}(s,t) =-{ \epsilon^{\mu\nu}\over i}
\int {d^2q\over 2\pi} {\rm tr}\left\{ \partial_\mu S_F^{-1}(q)\partial_\nu 
S_F(q)\right\}_{ss}
	\nonumber \\
& = \displaystyle {\epsilon^{\mu\nu}\over i} \left[ \int {dq_\mu \over 2\pi}
{\rm tr}\left\{ \partial_\mu S_F^{-1}(q)\cdot S_F(q)\right\}_{ss}
\right]_{q_\nu =\epsilon}^\pi \nonumber
\end{eqnarray}
This would be zero if there were
no infra-red singularities in $S_F$.
However, because of the contribution from zero modes,
$S_F$ is singular at $q=0$. Therefore
\widetext
\begin{eqnarray}
\sum_t \Gamma_{CS}(s,t) & = & - \sum_t\Gamma_{CS}(t,s) 
  =  \displaystyle
4 \lim_{\epsilon\rightarrow 0} \left[ \int_{\epsilon}^{\pi/2} 
{dq_\mu\over 2\pi}\sum_X \delta_X
[S_\nu (q) C_\mu (q) G_X^0 (q)]_{ss}\right]_{q_\nu =\epsilon}^{\pi /2}
	\nonumber \\
&=&  \displaystyle
-4 \sum_X \delta_X F_X(s,s)
\lim_{\epsilon\rightarrow 0} \int_{\epsilon}^{\pi/2} {dq_\mu\over 2\pi}
{\epsilon\over q_\mu^2 + \epsilon^2}
	\nonumber \\
& = &  \displaystyle
-\sum_X 2\delta_X
{F_X(s,s)\over\pi} \left[\tan^{-1}{q\over\epsilon}\right]_{\epsilon}^{\pi/2} 
        \nonumber \\
& = & \displaystyle
-\sum_X {\delta_X\over 2} F_X(s,s) ,
\label{sum}
\end{eqnarray}
\narrowtext where $F_X(s,t)$ is given in the previous subsection.

Since $S_F$ becomes the Wilson fermion propagator with constant mass term
for $1 \ll |s|,|t|, |L-s|, |L-t|$ with $s\cdot t > 0$ [see eq.(\ref{wil}) ],
it becomes
\begin{eqnarray} & 
\sum_{s,t}\Gamma_{CS}^{\mu\nu}(s,t)A_\mu(s,p) A_\mu(t,-p)
\longrightarrow
	\nonumber \\
& -\epsilon^{\mu\nu} \int dp_3 A_\mu(p_3,p)p_3 A_\mu (-p_3,-p)
\displaystyle \int {d^3 q\over (2\pi)^2}
	\nonumber \\
& \times
{\rm tr}\left\{ [\partial_\mu S_F^{-1}\cdot S_F ]
[\partial_3 S_F^{-1}\cdot S_F ] [\partial_\nu S_F^{-1}\cdot S_F ]
\right\} ,
\label{cst}
\end{eqnarray}
which coincides with the result of ref.\cite{CS}. This is a good check of our 
calculation.  From Ref.\ \cite{CS} we obtain
\begin{equation}
= \epsilon^{\mu\nu} \int dp_3 A_\mu(p_3,p)p_3 A_\nu (-p_3,-p)
\times \left\{
\begin{array}{ll}
\displaystyle 1 & \mbox{\quad for $+m_0$} \\
	& \\
0 & \mbox{\quad for $-m_0$}
\end{array} .
\right.  \nonumber
\end{equation}

The parity-even term $K$ satisfies
$K(s,t)=K(t,s)$ and
\begin{eqnarray}
 & \sum_{t}& K(s,t) =  \sum_t K(t,s) 
	\nonumber \\
& = & \displaystyle
- \int  {d^2 q\over (2\pi)^2}
{\rm tr}\left\{
[\partial_\mu S_F^{-1}\cdot\partial_\mu S_F]_{ss} + [\partial_\mu^2 S_F^{-1}
\cdot S_F ]_{ss} \right\}
        \nonumber \\
 & = & \sum_X {1\over 2}\cdot F_X (s,s)
\label{sumk}
\end{eqnarray}
The derivation of the last equality is similar to that of eq.(\ref{sum}).

\subsection{Total Contributions} 
Combining the above contributions  we finally obtain
\widetext
\begin{eqnarray}
S_{eff}^{(2)} &=& 
-{g^2\over 4\pi} \sum_{s,t} \int d^2 x \left\{ 
\sum_X F_X(s,t)^2
[A_\mu(s,x)(\delta^{\mu\nu}-{\partial_\mu\partial_\nu \over \Box}) A_\nu (t,x)]
         \right.
         \nonumber \\
&+&
[K(s,t)- \sum_X {1\over 2}\cdot F_X(s,t)^2 ] A_\mu (s,x) A_\mu (t,x)
         \nonumber \\
 &+& \left.
\sum_X i\delta_X F_X(s,t)^2   [{\partial_\mu\over \Box }
A_\mu (s,x)\epsilon^{\alpha\nu}\partial_\alpha A_\nu (t,x) ]+  
i\Gamma_{CS}(s,t) \epsilon^{\mu\nu} A_\mu(s,x) A_\nu(t,x) 
\right\}  .
\label{eff}
\end{eqnarray}
\narrowtext
This is the main result of this paper.
It is noted that the above formula is valid for both the Kaplan's and
the Shamir's methods.
The following consequences can be drawn from eq.\ (\ref{eff}) above.

The Parity-odd terms, which are proportional to $\epsilon^{\alpha\nu}$,
are unambiguously defined, contrary to the case of the continuum regularization
for anomaly free chiral gauge theories\cite{FS} 
which only regulates the parity even terms\cite{Neu,AK}.
These parity-odd terms break gauge invariance in the 2-dimensional sense.

For {\it anomalous} chiral  Schwinger model,
the parity-odd term with $X=R$ is localized around $s=0$ 
and that for $X=L$ is localized around $s=L$.
The effective action above for anomalous chiral Schwinger model via
the Kaplan's method or the Shamir's variation is different from the one
via the usual  Wilson fermion in 2 dimensions\cite{chiral}:
The term proportional to $\Gamma_{CS}$, which can not be evaluated analytically
for $s$ dependent gauge fields, is special for chiral fermions
from 3-dimensional theories, and the presence of this term prevents us from
concluding whether the anomalous chiral Schwinger model can be consistently
defined via the Kaplan's (Shamir's) method.

For anomaly free cases such that $\sum_R g_R^2 = \sum_L g_L^2$,
the parity-odd terms are exactly cancelled {\it locally} in $s$ space.
Here $g_{R(L)}$ is the coupling constant of a fermion with positive 
(negative) $m_0$ which generate a right-handed (left-handed) zero mode
around $s=0$.  The simplest but non-trivial example is
a Pythagorean case, $g_R=3, 4$ and $g_L = 5$\cite{Dis}.
Even for these anomaly free cases, 
the longitudinal term, whose coefficient is $K-F^2/2$,
remains non-zero in the effective action,
so that gauge invariance in the {\it 2-dimensional} sense is
violated.
In this regard
the form of the effective action via the Kaplan's
(Shamir's) method is similar to
the one via the usual  Wilson fermion\cite{chiral}.

Let us consider the effective action for $s$ independent gauge fields
as in ref.\cite{Neu}.
Since $\sum_{\mu ,\nu }\epsilon^{\mu\nu} A_\mu(x) A_\nu(x) = 0$
for $s$ independent gauge fields, the Chern-Simons term vanishes.
The other parity-odd term is cancelled between the two zero modes
since $\sum_{X,s,t} \delta_X F_X(s,t)^2=0$.
The longitudinal term also vanishes due to the identity
\begin{equation}
\sum_{t}[K(s,t)-{F_X(s,t)^2\over 2}] = 
\sum_{t}[{F_X(s,t)^2\over 2}-{F_X(s,t)^2\over 2}] = 0
\end{equation}
[see eqs.\ (\ref{sumf}) and (\ref{sumk})].
Therefore the effective action becomes
\begin{equation}
S_{eff}^{(2)} =
-2\times {g^2\over 4\pi}  \int d^2 x 
[ A_\mu(x)(\delta^{\mu\nu}-{\partial_\mu\partial_\nu \over \Box}) A_\nu (x)].
\end{equation}
This effective action is transverse and thus gauge invariant 
in the 2-dimensional sense.  Both zero modes around $s=0$ and $s=L$
equally contribute so that a factor 2 appears in the above result.
This is consistent with the general formula derived in 
ref.\cite{Neu2}.   
The anomalous chiral Schwinger model can not be simulated
by the Kaplan's (Shamir's) method with the $s$-independent gauge fields,
since the gauge fields see {\it both} of the zero modes so that 
it fails to reproduce the parity-odd term, expected to exist\cite{JR}

\section{Anomalies in the Chiral Schwinger Model}
\label{sec:anomaly}

\subsection{Currents and their Divergence}

From the effective action obtained in the previous section, we
can calculate the vacuum expectation values of various currents
in the presence of back-ground gauge fields.
Let us define the fermion number current as 
\begin{equation}
\langle J_\mu^g (s, x) \rangle
=  {\delta S_{eff}^{(2)} \over g\delta  A_\mu (s, x) }
\end{equation}
where the index $g$ in the current explicitly shows the charge of the
fermion. 
From eq. (\ref{eff}) we obtain
\widetext
\begin{eqnarray}
 J_\mu^g(s,x)  &=& i{g\over 4\pi}\sum_t
 [ \sum_X \delta_X F_X(s,t)^2(\epsilon^{\alpha\nu}\partial_\mu
+\epsilon^{\alpha\mu}\partial_\nu ){\partial_\alpha\over \Box}A_\nu (t, x)
-2\Gamma_{CS}\epsilon^{\mu\nu} A_\nu (t,x)] 
	\nonumber \\
&-& {g\over 4\pi}\sum_t
[2\sum_X F_X(s,t)^2
(\delta^{\mu\nu}-{\partial_\mu\partial_\nu \over \Box}) A_\nu (t,x)
+
(2 K(s,t)- \sum_X F_X(s,t)^2 )  A_\mu (t,x) ] 
	\nonumber \\
& \equiv &  J_\mu^{g,odd}+ J_\mu^{g,even}
\label{current}
\end{eqnarray}
\narrowtext
where $J_\mu^{g,odd}$ is a parity-odd current (the first two 
terms) and $J_\mu^{g,even}$ is a parity-even current (the last two
terms). Hereafter 
all $J_\mu$ should be understood as vacuum expectation values,
though $\langle \qquad\rangle$ is suppressed.
From the fermion number current
the gauge current for a fermion with charge $g$ is easily constructed as
$ J_\mu^G (s,x) \equiv g J_\mu^g (s,x)$. 

Divergences of the parity-odd and parity-even currents become
\begin{eqnarray}
\partial_\mu J_\mu^{g,odd}(s,x)
 &= & i {g\over 4\pi} \sum_t[\sum_X \delta_XF_X(s,t)^2- 2\Gamma_{CS}(s,t)]
	\nonumber \\
&\times &  \epsilon^{\mu\nu} \partial_\mu A_\nu (t,x)  ,
\end{eqnarray}
\begin{eqnarray}
\partial_\mu J^\mu_{g,even}(s,x)
 & = &  {g\over 4\pi} \sum_t [\sum_X F_X(s,t)^2- 2K(s,t)] 
	\nonumber \\ 
&\times & \partial_\mu A_\mu (t,x)  .
\end{eqnarray}

\subsection{Gauge invariance}

As mentioned in Sect.\ \ref{sec:model}, 
the action in $A_3 = 0$ gauge is invariant
under $s$ independent gauge transformation. This invariance implies
$\sum_{s} \partial_\mu J_\mu^G (s,x) = 0$. This identity is satisfied in 
our calculation of the effective action, since
\begin{eqnarray}
 \sum_s & [\sum_X\delta_XF_X(s,t)^2-  2\Gamma_{CS}(s,t)]
	\nonumber \\
&= \sum_X \delta_X [F_X(t,t)-F_X(t,t)] =0 
	 \label{sum0a}
\end{eqnarray}
\begin{eqnarray}
 \sum_s & [\sum_X F_X(s,t)^2- 2 K(s,t)] 
	\nonumber \\
&= \sum_X [F_X(t,t)-F_X(t,t)] =0  
\label{sum0b}
\end{eqnarray}
from eqs.(\ref{sumf},\ref{sum},\ref{sumk}).

\subsection{Gauge Anomalies}
The gauge anomaly for a fermion with a charge $g$, denoted $T_g$, is defined by
$ T_g = g\partial_\mu J_\mu^{g,odd}$, and it becomes
\begin{equation}
T_g (s,x) = \sum_t g^2 C(s,t)\times T^0(t,x) 
\end{equation}
where 
\begin{equation}
T^0(t,x) =
i{1\over 4\pi} \epsilon^{\mu\nu} \partial_\mu A_\nu (t,x)
\end{equation}
is the gauge anomaly of a {\it 2-dimensional} theory, and
\begin{equation}
C(s,t) = \sum_X \delta_XF_X(s,t)^2-2 \Gamma_{CS}(s,t)
\end{equation}
represents the spread of the gauge anomaly over the 3rd direction
due to the spread of zero modes.
This spread of the anomaly has been observed in a numerical 
computation\cite{Jan}.
It is noted that the divergence of the gauge current $J_\mu^G$ also contains
parity-even contributions, given by
\begin{equation}
\sum_t D(s,t)\times {g^2\over 4\pi} \partial_\mu A_\mu(t,x)
\end{equation}
where $ D(s,t) = \sum_X F_X(s,t)^2-2 K(s,t)$.

The one-loop integral (\ref{gcs}) defining
$\Gamma_{CS}$ is 
too complicated to calculate analytically.
For $t$-independent gauge fields $A_\mu (t,x) = A_\mu (x)$
there is a considerable simplification and
we obtain
\begin{equation}
T_g (s,x) =   g^2C(s)\times T^0(x) 
\end{equation}
where
\begin{equation}
T^0(x)= i{1\over 4\pi} \epsilon^{\mu\nu} \partial_\mu A_\nu (x)
\end{equation}
and
\widetext
\begin{eqnarray}
C(s)  &=& \sum_t C(s,t) = 2 \sum_X \delta_X F_X(s,s)  
	\nonumber \\
&=& {m_0(4-m_0^2)\over 2}
\times  \left\{
\begin{array}{ll}
\left[ (1-m_0)^{2s}-(1-m_0)^{2(L-s-1)} \right] & \mbox{\ for $s \ge 0$} \\
	& \\
\left[ (1+m_0)^{2s}-(1+m_0)^{-2(L+s+1)}\right] & \mbox{\ for $s \le 0$}
\end{array}  
\right. .
\end{eqnarray}
\narrowtext
for the Kaplan's method. We plot $C(s)$ as a function of $s$ 
at $m_0= 0.1$ and 0.5 in Fig.\ref{anomalyK}. 
For the Shamir's method, we obtain
\begin{equation}
C(s) = 2m_0(2-m_0)\cdot
\left[ (1-m_0)^{2s}-(1-m_0)^{2(L-s)} \right]
\end{equation}
and plot this in Fig.\ref{anomalyS}.
It is noted that there is no parity-even contribution in $\partial_\mu J_\mu^G$
since $\sum_t D(s,t) = 0$ in this case.

\subsection{Chern-Simons Current}
From the 3-dimensional point of view, the gauge anomaly should be cancelled
in such a way that $T_g 
+ \partial_3 gJ_3^{CS}(s,x)=0$\cite{KP,CS}, where $J_3^{CS}$ is the
third component of the Chern-Simons current for the 3-dimensional 
vector gauge theory. With our gauge fixing the effective action does not
depends on $A_3$, and  it is difficult to calculate $J_3^{CS}$ analytically
except in the region away from domain-walls\cite{CS}.
However, for $t$-independent gauge fields,
we can obtain the Chern-Simons current everywhere via the relation
$\partial_3 gJ^3_{g,odd}(s,x) = -T_g$, which becomes
\begin{equation}
J_3^{CS}(s+{1\over 2},x)-J_3^{CS}(s-{1\over 2},x) = g^2C(s)\times T^0 (x).
\end{equation}
Taking $J_3^{CS}(s+{1\over 2},x)= g^2 I(s)\cdot T^0 (x)$, we obtain 
\begin{equation}
I(s+{1\over 2})-I(s-{1\over 2}) = C(s)  .
\end{equation}
We have to solve this equation with the boundary condition
$ I(s)\rightarrow -2$ as $s\rightarrow +\infty$\cite{CS}.
For a finite $s$ space $s\rightarrow +\infty$ means
$ 1\ll s \ll L$.

For the Kaplan's method we obtain
\widetext
\begin{equation}
I(s-{1\over 2})=\left\{
\begin{array}{ll}
\displaystyle
{2+m_0\over 2}\cdot [(1-m_0)^{2s}+(1-m_0)^{2(L-s)}]-2 & \mbox{\quad 
$0\le s\le L$} \\
	& \\
\displaystyle -{2-m_0\over 2}\cdot 
[(1+m_0)^{2s}+(1+m_0)^{-2(L+s)}] & \mbox{\quad $ -L\le s\le 0$}
\end{array}
\right. .
\end{equation}
\narrowtext
This solution automatically satisfies the other boundary condition
that $ I(s)\rightarrow 0$ as $\rightarrow -\infty$\cite{CS}.
Again $\rightarrow -\infty$ means $ 1\ll -s \ll L$ for a finite $s$ space.
We plot $I(s)$ as a function of $s$ at  $m_0=0.1$ and 0.5 in Fig.\ \ref{CSK}.
For the Shamir's method we obtain
\begin{eqnarray}
I(s-{1\over 2}) &= 2[(1-m_0)^{2s}+(1-m_0)^{2(L-s+1)}]-2 
	\nonumber \\
 & \qquad \qquad \mbox{for $0\le s\le L-1$},
\end{eqnarray}
which is plotted in Fig.\ \ref{CSS}.

\subsection{Pythagorean Chiral Schwinger Model and
Anomaly in Fermion Number Current}

Let us consider the Pythagorean chiral Schwinger model\cite{Dis}.
In this model there are 
two right-handed fermions with charges $g_1$ and $g_2$, and
one left-handed fermion with charge $g_3$.
Formulation of this model via the Kaplan's method has already been discussed
in ref.\cite{KP,Jan,Dis}
( an extension to the Shamir's method is straightforward ).
We assign $+m_0$ for fermions with charge $g_1$ and $g_2$, and
$-m_0$ for a fermion with charge $g_3$. The value of  $|m_0|$ should
be equal for all fermions, as will be seen below.

The theory has a $U(1)^3$ symmetry\cite{Dis} corresponding to independent phase
rotations of three fermions. The corresponding currents are
\begin{equation}
\left\{
\begin{array}{lll}
J_\mu^G &=& g_1 J_\mu^{g_1}+g_2 J_\mu^{g_2}+g_3 J_\mu^{g_3} \\
        & & \\
J_\mu^R &=& g_2 J_\mu^{g_1}-g_1 J_\mu^{g_2} \\
        & & \\
J_\mu^F &=&  J_\mu^{g_1}+J_\mu^{g_2}+J_\mu^{g_3}  
\end{array}
\right.
\end{equation}
The first one is the gauge current, whose divergence becomes
\begin{equation}
\partial_\mu J_\mu^G(s,x) = (g_1^2+g_2^2-g_3^2)\times \sum_t C(s,t)\cdot
 T^0(t,x).
\end{equation}
Therefore, if $g_1^2+g_2^2 = g_3^2$ (Pythagorean relation) is satisfied
and if all fermions
have the same value of $|m_0|$ to give the same $C(s,t)$,
this current is conserved and there is no gauge anomaly for any
background gauge fields. 
The second current is non-anomalous, since
\begin{equation}
\partial_\mu J_\mu^R (s, x)=(g_2\cdot g_1 - g_1\cdot g_2)\sum_t 
C(s,t)\cdot T^0(t,x) = 0  .
\end{equation}

The third current, which corresponds to the fermion number of the theory,
is anomalous, since
\begin{equation}
\partial_\mu J_\mu^F (s, x)=( g_1 + g_2 - g_3)\times\sum_t C(s,t) \cdot
T^0(t,x)   .
\end{equation}
The Kaplan's method as well as the Shamir's one 
successfully give a non-zero divergence for the fermion number current, though
the coefficient $C(s,t)$ has a finite width.
For $t$-independent gauge fields, this anomaly becomes
\begin{equation}
( g_1 + g_2 - g_3)\cdot C(s)\cdot  T^0(x)
\end{equation}
where $C(s)$ is almost localized at $s=0$ and at $s=L$ as seen in
Fig.\ref{anomalyK} and Fig.\ref{anomalyS}.
Since the fermion number is conserved in the 3-dimensional
theory,  the third component of the fermion number current should satisfy
$\partial_3 J_3^F + ( g_1 + g_2 - g_3)\cdot C(s)\cdot T^0(x) = 0$\cite{Dis}. 
Therefore we obtain
\begin{equation}
J_3^F(s,x)= (g_1+g_2-g_3)\cdot I(s)\cdot T^0(x)  .
\end{equation}

\section{conclusions}
\label{sec:concl}
In this paper we have formulated a lattice perturbative expansion for
the Kaplan's chiral fermion theories, extending the suggestion
by Narayanan and Neuberger\cite{Neu}. 
Applying our perturbative technique to the chiral Schwinger model
formulated via the Kaplan's or the Shamir's method,
we have calculates the fermion one-loop effective action for gauge fields.
The effective action contains parity-odd terms and
longitudinal terms, both of which break 2-dimensional gauge invariance,
and the anomaly of the gauge current is obtained from the effective action.
The gauge anomaly is calculable in the Kaplan's (Shamir's)
method if the perturbative expansion is carefully formulated.
For the anomaly-free Pythagorean chiral Schwinger model,
the fermion number current is anomalous. To obtain this anomaly
the fermion number current should not be summed over $s$, in contrast
to the case of the continuum calculation\cite{AK}, where
the anomaly comes from an infinite summation over $s$.

The main conclusions drawn from the results are as follows.

\begin{enumerate}
\item 
Anomaly of the fermion number current is shown to be
non-zero in this method, though the current flows off walls
into the extra dimension.
Since the current is external we feel that this
does not affect the dynamics of the model
and therefore does not spoil the 2-dimensional nature of the chiral
zero mode on the perturbtive level.
It was argued in ref.\ \cite{Dis}, however, that
fermionic non-zero modes in the trivial background of gauge fields
may become zero modes in an instanton background
so that the theory ceases to be 2 dimensional on the non-perturbative level.
Whether this is true or not is cruicial for the success of
the Kaplan's (Shamir's) method,
and therefore this question should be investigated by the numerical method.

\item Two-dimensional gauge invariance at low energy
can not be assured by the Kaplan's (Shamir's) method, 
except for $s$-independent
gauge fields, even for anomaly-free cases.
This is similar to the situation with
lattice chiral gauge theories formulated with the ordinary Wilson mass 
term\cite{chiral}.
In this point the Kaplan's (Shamir's) method does not seem  better than the 
conventional approaches.
At this moment it is not clear whether this violation of gauge invariance
spoils the whole program of this method. In particular the effects of the 
longitudinal component of gauge fields has to be analyzed further.

\item If the theory is anomaly free and  gauge fields are
$s$-independent\cite{Neu}, the gauge invariance as a 2-dimensional theory
can be maintained. However, the gauge fields feel both of the zero modes 
even in the $L\rightarrow \infty$ limit, and the fermion loop contribution
to the effective action is twice as large as the one expected from a single
chiral fermion. Therefore we have to take a square-root of 
the fermion determinant
to obtain the correct contribution.  
For fermion quantities such as the fermion number current, however,
it seems possible
to separate the contribution of the chiral zero mode at $s=0$ from
that of the anti-chiral zero mode at $s=L$, as seen in the previous section.
\end{enumerate}

Perturbative calculations performed in 
this paper can  be extended to 4+1 dimensional theories. Of course
actual calculations become much more complicated and difficult
because of severe ultra-violet divergences in 4+1 dimensions than 
in 2+1 dimensions. 
Work in this direction is in progress. 

\acknowledgements
We would like to think Prof. Ukawa for discussions and the careful reading of
the manuscript.

After finishing this work, a new paper by Narayanan and Neuberger\cite{Neu3}
appeared. In the paper
the gauge anomaly for the chiral Schwinger model was calculated
semi-analytically via the overlap formula of ref.\cite{Neu2}.

\newpage

\begin{figure}
\caption{ Two zero modes $u_L$ and $u_R$
as a function of $s$ at 
$m_0=0.1$ and 0.5 for $p_1=p_2=0$. We take $L=100$.}
\label{zero}
\end{figure}

\begin{figure}
\caption{ The coefficient of the anomaly $C(s)$ 
for the Kaplan's method as a function of $s$ 
at $m_0=0.1$ and 0.5 for $L=100$.}
\label{anomalyK}
\end{figure}

\begin{figure}
\caption{ The coefficient of the anomaly $C(s)$ 
for the Shamir's method
as a function of $s$ at $m_0=0.1$ and 0.5 for $L=100$.}
\label{anomalyS}
\end{figure}

\begin{figure}
\caption{The coefficient of the Chern-Simons current $I(s)$
for the Kaplan's method as a function of $s$ 
at $m_0=0.1$ and 0.5 for $L=100$.}
\label{CSK}
\end{figure}

\begin{figure}
\caption{ The coefficient of the Chern-Simons current $I(s)$ 
for the Shamir's method as a function of $s$ 
at $m_0=0.1$ and 0.5 for $L=100$.}
\label{CSS}
\end{figure}

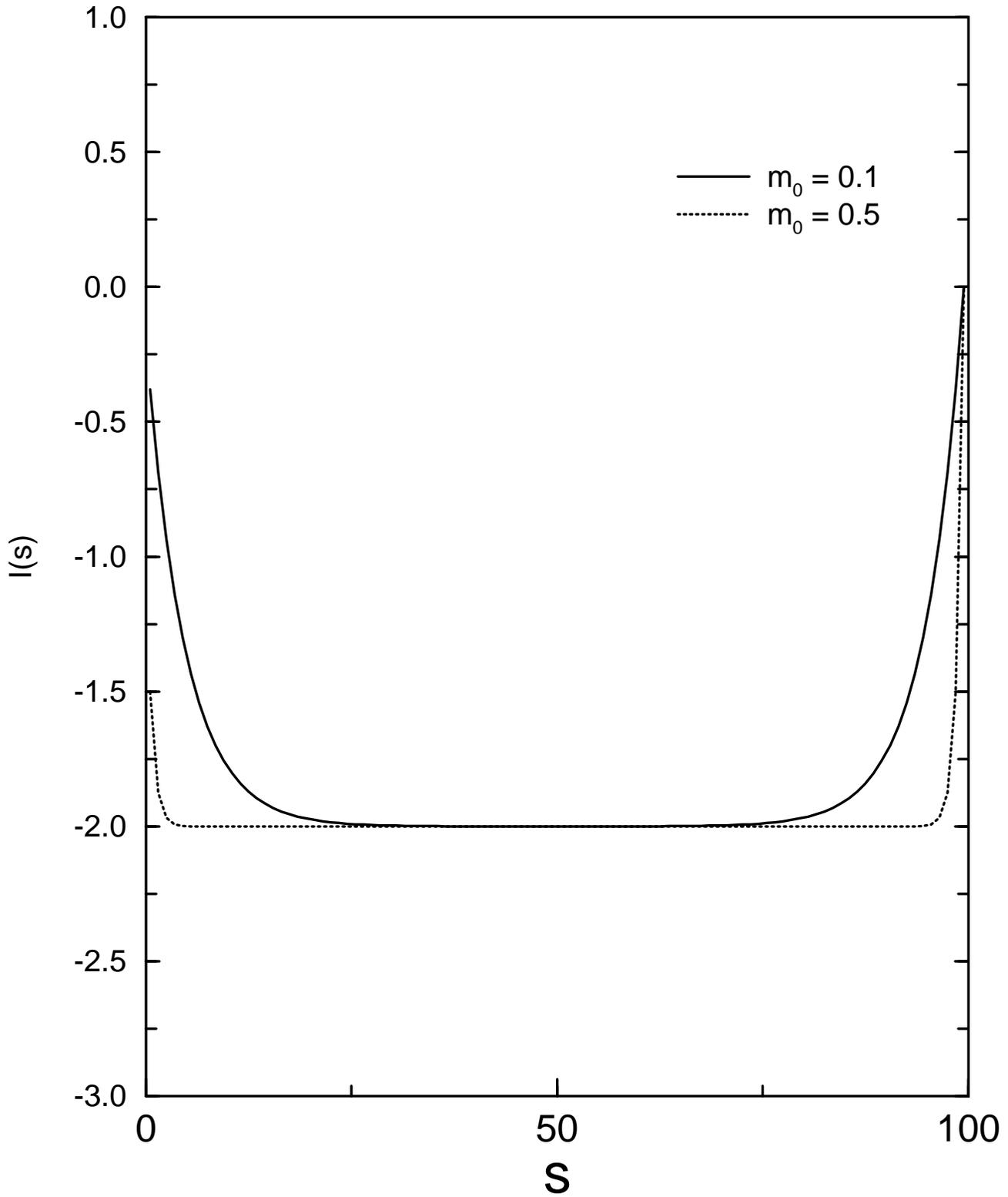
\end{document}